# MedGraph: An experimental semantic information retrieval method using knowledge graph embedding for the biomedical citations indexed in PubMed

Islam Akef Ebeid
Department of Information Science
University of Arkansas at Little Rock
Little Rock, Arkansas, USA
iaebeid@ualr.edu

Elizabeth Pierce
Department of Information Science
University of Arkansas at Little Rock
Little Rock, Arkansas, USA
expierce@ualr.edu

## Abstract

Here we study the semantic search and retrieval problem in biomedical digital libraries. First, we introduce MedGraph, a knowledge graph embedding-based method that provides semantic relevance retrieval and ranking for the biomedical literature indexed in PubMed. Second, we evaluate our method using PubMed's Best Match algorithm. Moreover, we compare our method MedGraph to a traditional TFIDF based algorithm. We use a dataset extracted from PubMed, including 30 million articles' metadata such as abstracts, author information, citation information, and extracted biological entity mentions. We do that by pulling a subset of the dataset to evaluate MedGraph using predefined queries with ground truth ranked results. To our knowledge, this technique has not been explored before in biomedical information retrieval. In addition, our results provide evidence that semantic approaches to search and relevance in biomedical digital libraries that rely on knowledge graph modeling offer better search relevance results when compared with traditional approaches in terms of objective metrics.

## 1 Introduction

PubMed is the National Library of Medicine's (NLM) free authoritative database of citations and search engine of more than 30 million articles in biology, medicine, pharmacy, and life sciences and across multiple curated databases such as MEDLINE (http://www.nlm.nih.gov/pubs/factsheets/medline.html). PubMed is used by more than 2.5 million users each day, serving clinicians, physicians, researchers, and students (Fiorini et al., 2018). It is worth mentioning that PubMed is a database of citations, not a database of full-text articles. That is because about two-thirds of the articles indexed in PubMed do not provide access to full texts (https://pubmed.ncbi.nlm.nih.gov/). Instead, when a free full text is available by the publisher or published as open access or is supported by an NIH grant, the full article gets indexed in PubMed Central, NLM's accessible repository of full-text articles. Accordingly, the PubMed search engine relies on metadata and citations instead of parsing full-text articles when providing a search experience. Articles' metadata are indexed and parsed in fields to be utilized in the search process. Metadata fields include titles, abstracts, authors, journal names, publication dates, submission dates, related MeSH terms, citation and references information, funding grants, projects, and many more.

PubMed uses an algorithm that relies on fuzzy string matching to match the query with relevant citations. For example, when a user enters in the search box an author name followed by a journal name, all the articles that that author published in that journal will appear. In addition, PubMed uses the Automatic Term Mapping system (ATM) (Adlassnig & others, 2009). The ATM system expands the input query and finds which fields the query entered intended. The expanded query is then matched with the most relevant documents using MeSH terms, keywords, and other metadata that could be treated as an index. The most relevant articles are then retrieved using TF-IDF and ranked based on date or alphabetically using either the title or the author name (Fiorini et al., 2018).

Other methods include ranking by date or author information. Recently PubMed deployed its most recent relevance ranking algorithm named BestMatch (Fiorini et al., 2018). BestMatch relies on a machine learning model trained on features extracted from user search logs on PubMed in the past several years. The system has been shown to outperform TFIDF based ranking. However, BestMatch does not consider that the user query logs that the system has been trained on contain ambiguous queries. In addition, even though the authors evaluated the system thoroughly using an A/B testing approach with real users to evaluate the ranking quality, the algorithm did not provide solutions for the problem of understanding query intentions through semantic models. The intention of the query is fundamental.

For example, a user can enter the word "cancer" in the PubMed search box, and they might mean multiple things by "cancer." For example, they might want an article in the journal named "Nature: Cancer." Alternatively, they might want authors who work and publish in the field of cancer. Alternatively, they might want all relevant articles that mention cancer or research done in the field of cancer. They might also be looking for a specific citation with a title or author name, journal, and year. Alternatively, they might be looking for several citations. Search engines and information retrieval systems such as PubMed and Google rely on objective metrics and algorithms to rank their search results. The ranking of the search results does not necessarily reflect what the user meant by the query. They, however, reflect the most objective relevance based on the text of the input query. That is done by analyzing the frequency of the strings in the input queries in the corpus of documents. In addition, other models incorporate the citation network of the documents, such as PageRank in the case of Google (Page et al., 1999). Hence, incorporating semantics in search algorithms and information retrieval systems, especially in biomedical literature searches, is crucial to move towards systems that can sort out ambiguity, understand query intentions, and aid in true knowledge discovery. In recent years and with the Web 2.0 information revolution, Semantic Web technologies have proliferated (Berners-Lee et al., 2001). Semantic web technologies aim to create an understandable and readable web by machines. The graph model was introduced to represent knowledge in web pages semantically using standards such as the Resource Descriptor Framework (Lassila et al., 1998). The idea was driven by earlier work in digital ontology and concept maps. Knowledge graphs were then born as a data model used to store information and data semantically. Knowledge graphs have also been extended as graph databases for data persistence as it allows for a more flexible representation of data and relationships when compared with the relational data model (Hogan et al., 2020).

To overcome the problems associated with semantic understanding of queries when searching the biomedical literature in PubMed, we introduce MedGraph, a knowledge graph-based search engine and information retrieval method. MedGraph relies on converting the metadata associated with PubMed into a knowledge graph. The metadata includes disambiguated author names, grant information, MeSH terms, citation information, and a dataset of extracted bio entities such as drugs, genes, proteins, species from the text of the title and the abstract of each article in PubMed. The dataset was introduced in (Xu et al., 2020), and it includes NIH project involvement for each author and each article in PubMed. In addition, it includes extracted biological entities using deep learning named entity recognition technique named BioBERT (Lee et al., 2020). The dataset is available as a relational database linked using each article's unique identifier PMID. The dataset contains articles from the year 1781 until December 2020. To provide a proof of concept for the utility of MedGraph, we extracted a small dataset of 2696 articles and their associated metadata and citation network from the PubMed dataset (Xu et al., 2020). We then extracted the entities from the dataset and linked them semantically as a knowledge graph. We then used a knowledge graph embedding method named Node2vec (Grover & Leskovec, 2016) to vectorize and embed the extracted knowledge graph in a Euclidean space. We then used the node vectors to rank the articles using a cosine distance similarity measure on the learned vectors according to the input query after pooling all the vectors of related first-order neighbor nodes for each article. On the query side, first, the input query is parsed and expanded using the extracted biological entities in the original dataset as an index. The expanded query is then matched to their corresponding nodes in the knowledge graph. The matched node vectors are then averaged to vectorize each query. We perform a complete evaluation on the proposed method against PubMed's BestMatch algorithm as ground truth using various metrics. In addition, we compare our method with a traditional TF-IDF approach (Ramos & others, 2003) (Jones, 1972). Our results show that MedGraph performs comparably to BestMatch. In addition, it outperforms the traditional TF-IDF method providing evidence that using knowledge graph-based semantic search will benefit the biomedical and life science research community when adopted as a widely used method in literature search through digital libraries.

In the next section, 2, we give a background on knowledge graphs and review the literature on previous work in biomedical-based information retrieval using graph technologies. In section 3, we describe our method and the MedGraph framework. In section 4, we describe our evaluation experiments and results. Section 5 discusses the results, implications, and future work and concludes in section 6. A complete bibliography is available in section 7.

## 2 Background
### 2.1 Knowledge graphs
The term "Knowledge Graph" has emerged to describe a new Google search technology in 2012 (Paulheim, 2016). The term has expanded to describe any form of a knowledge base represented using a graph structure (Ji et al., 2020). Knowledge graphs are data models used to represent structured and unstructured knowledge in triples <Mount Fuji, isLocatedIn, Japan> (Ji et al., 2020). The triples represent a semantic relationship between two objects following language semantics written in first-order logic such as <subject, verb, object>. Hence, a knowledge graph represents interrelated and semantically connected descriptions of real-world entities and relationships (Ji et al., 2020). The definitions of a knowledge graph vary widely (Li et al., 2020). Here I assume the definition described in (Paulheim 2016). In Paulheim et al. (2016), the authors identified a set of properties that pertains to a knowledge graph that is:

- They describe real-world phenomena in the form of entities and relationships.

- The entities and relationships are represented in a graph structure using vertices and edges.

- The graph structure containing the knowledge graph can be described mathematically and represented computationally using graph theory norms.

- The entities' metadata and relationships do not necessarily adhere to a schema or an ontology.

- Entities of different types could be linked with multiple edges.

- It could cover multiple domains of knowledge.

Knowledge graph construction can vary from manual to automatic curation by extracting entities and relations from unstructured text (Paulheim, 2016). Knowledge graphs can be represented using the property graph model (Angles, 2018). Knowledge graphs can also be represented using the Resources Descriptors Framework (Lassila et al., 1998), a semantic representation language with a graph data model defined by the World Wide Web Consortium. A relationship is defined between two connected entities or nodes, such as our previous example <Mount Fuji, isLocatedIn, Japan>. In that example <Mount Fuji> and <Japan> are two unique nodes, vertices or entities in a graph, while <isLocatedIn> is the relationship or edge label connecting them. The RDF standard requires storing node names using a Unique Resource Identifier (URI) format. Knowledge graphs can be constructed and curated under two assumptions: the open-world assumption or the closed world assumption (Cai et al., 2018). In the open-world assumption, information outside the knowledge graph is assumed to be unknown, and the knowledge graph itself is considered incomplete, such as DBpedia (Bizer et al., 2009). In a closed world assumption, information outside the knowledge graph is false. The data model in the closed-world assumption can be defined in a schema or ontology described using the RDFS language or the Resource Descriptor Framework

Schema (Lassila et al., 1998). A schema is a concept map or ontology describing entities in the knowledge graph and their relationships.

Like relational databases, knowledge graphs are graph databases queried using query languages (Hogan et al., 2020). SPARQL (Harris, n.d.) and Cypher (*Cypher Query Language*, n.d.) are two popular graph query languages like the Structured Query Language (SQL). SPARQL was developed as part of the semantic web technology stack to query RDF triple stores and KGs, while Cypher is a property graph query language (Hogan et al., 2020). However, due to the unconventional structure of the graph model and the fact that knowledge graphs can be multigraphs with attributes and edges that could be weighted or labeled, graph query languages are difficult and impractical for the complex analysis of knowledge graphs. Those limitations have led to the emergence of knowledge graph mining methods and algorithms especially embedding models borrowing from graph theory, complex networks, machine learning, and deep learning to adapt approaches developed in those fields to mine knowledge graphs more flexibly and extract deeper insights (S. Wang et al., 2017).

## 2.2 Knowledge graph embedding and similarity

Knowledge graph embedding models automatically extract latent feature vectors from data without relying on stochastic and heuristic metrics and measures (Bengio et al., 2013). All representation learning algorithms on graphs ultimately produce node embedding vectors in low-dimensional vector space (Hamilton et al., 2018). The minimal constraint on the learned representations is that they preserve the graph's structure in the Euclidean vector space (Hamilton et al., 2018). The node embedding vectors learned can be passed to downstream machine learning models for classification, regression, or clustering. For example, cosine distance could be applied to the learned vectors to rank the nodes compared to a query node. Alternatively, they can be used directly in the learning process in a semi-supervised end-to-end fashion as in the Graph Neural Network (GNN) model (Wu et al., 2019). They can also be inductive as in learned from the structure of the graph itself (Perozzi et al., 2014) (Tang et al., 2015) (Grover & Leskovec, 2016) or transductive by forcing a scoring function to evaluate the plausibility of the triples in the knowledge graph (Lin et al., 2015).

The Skip-gram model described in (Mikolov, Chen, et al., 2013) is an unsupervised language model aiming at learning discrete vectors of unique words given a corpus of text. The Skip-gram model has been adapted to learn vectors for individual nodes on a corpus of sampled nodes from the graph in unsupervised network and graph embedding algorithms. In DeepWalk (Perozzi et al., 2014), the authors first proposed using the Skip-gram model on a corpus of sampled nodes from a given graph using a random walk strategy to sample the graph. Since it was observed that the distribution of words in any corpus of text followed Zipf's law, it was also observed that the frequency distribution of nodes in a semantic graph follows Zipf's law (Perozzi et al., 2014). The random walk algorithm aims at sampling chains of nodes from the graph controlled by the walk length. Sampled chains are then treated as sentences in a text corpus fed to a Skip-gram variant of Word2vec (Mikolov, Sutskever, et al., 2013).

## 2.3 Knowledge graphs in search engines, recommender systems, and biomedical digital libraries

Knowledge graphs have been adapted to aid search engines and recommender systems. They are highly efficient in those applications due to their flexibility in modeling the data. For example (Xiong et al., 2017), the authors introduced explicit semantic ranking harnessing the power of knowledge graph embedding. The algorithm uses graph representation learning on the metadata of articles in the online search engine named Semantic Scholar (Fricke, 2018). They use a knowledge graph embedding model to represent queries and documents as vectors in the same vector space. This work is the closest to the work we present here. The authors provided strong evidence that using knowledge graph embedding in searching academic literature improves the relevance of the returned documents drastically due to the reliance on semantics and entity matching in the process. While in (Q. Wang et al., 2017), the authors demonstrated the usefulness of knowledge graphs and semantic modeling in search engines when retrieving web pages. They used a relation extraction algorithm to construct a knowledge graph. Though they have not used graph embedding, they devised a semantic matching approach based on support vector machines.

In (Montes-y-Gómez et al., 2000), the authors introduced extracting a knowledge graph from the text of two documents. They then measured the similarity between these two graphs extracted from the two articles, combining relational and conceptual similarities. In (Ebeid et al., 2021), the authors showed the utility of ranking methods on embedded knowledge graphs using simple cosine distance metrics to perform tasks such as link prediction in the biomedical domain. While in (Matsuo et al., 2006), the authors described a system built using keyword co-occurrence matching. They remodeled the keyword matching process as a graph and applied a graph clustering technique to match keywords and queries. In (Blanco & Lioma, 2012), The authors modeled the text in documents as a graph instead of a Bag of Words model (BoW). Then, they used PageRank (Page et al., 1999) to derive similarity measures between documents. At the same time, the authors (Farouk et al., 2019) argued that graph modeling could enhance search relevance results based on context rather than just string similarity. They developed a system where the input documents and indices are converted to a knowledge graph. Their findings support that of (Ma et al., 2016) where they drove the point that graph-based search engines are highly efficient and valuable despite their challenges. Evidence on the utility of graph-based search is strengthened in (Guo et al., 2019), where the authors constructed a network of the standardized MeSH headings assigned to articles in MEDLINE (Motschall & Falck-Ytter, 2005). The relationships between the MeSH headings were modeled as a graph where the edges represent different hierarchical roles in the original MeSH coding system. The graph of MeSH headings was then fed to various graph embedding algorithms. The output was a learned feature vector representing each MeSH heading for each node. The data set is helpful in downstream biomedical computational tasks.

While in (J. Z. Wang et al., 2014), the authors used an efficient graph-based search engine on par with PubMed. Their approach tackled the problem of returning relevant documents from three angles. They first built a parallel document indexer. Second, they modeled each article's metadata, such as MeSH terms and keywords, as a graph and applied a personalized PageRank (Lofgren et al., 2016) to rank the concepts in the built graph, followed by TF-IDF (Pita et al., 2018) to rank the documents relative to a query. Third, they included the user's search behavior as a factor in relevance similar to BestMatch (Fiorini et al., 2018). Despite its efficiency compared to PubMed, the algorithm requires user input and is not fully unsupervised. The BestMatch (Fiorini et al., 2018) is the newest algorithm used by the PubMed search engine to find the most relevant articles to a user's query. BestMatch relies on extracting features from articles and including prior user search log into a relevance ranking prediction model. The model then finds the most relevant results personalized to each user. BestMatch provides excellent results compared with previous approaches in PubMed, yet it does not consider any semantics failing to distinguish ambiguity in queries.

## 3 Method

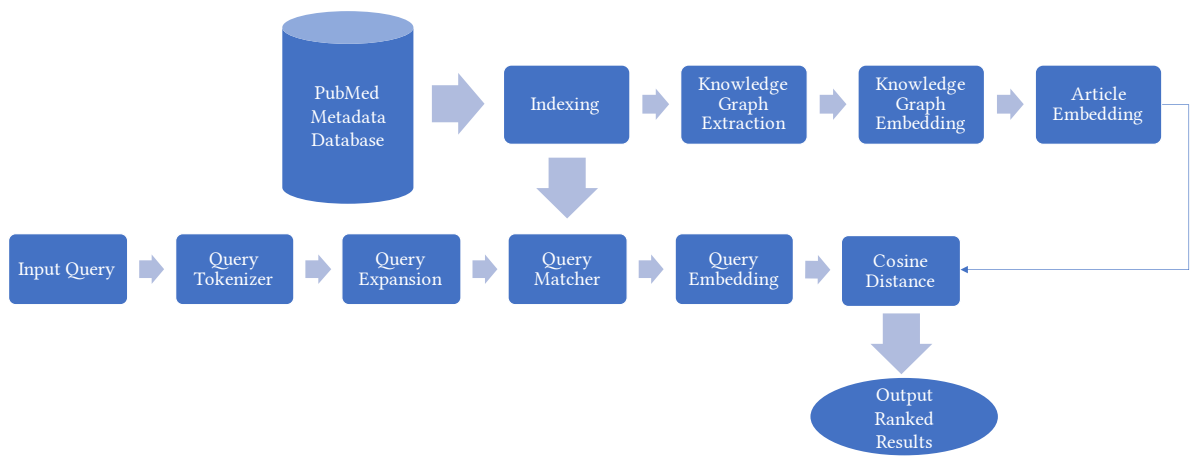

Figure 1. The MedGraph framework.

## 3.1 The PubMed metadata database

In (Xu et al., 2020), the authors extracted a metadata database from the corpus of the PubMed articles available from 1781 until December 2020 (30 million). The extracted information includes names of biological entities such as genes, proteins, species, drugs, diseases, in addition to disambiguated author information and citation information. The primary purpose of that dataset was to provide the basis for creating a full knowledge graph of the articles in PubMed. The extracted biomedical knowledge graph could be used in various biomedical information retrieval and data mining tasks. Here we utilize the extracted biomedical knowledge graph described in (Xu et al., 2020). The dataset comes as a relational database linked by a unique identifier, each article's identifier in PubMed, also known as the PMID. Those account for 31929000 articles. Author information from each article, including first names, middle names, last names, and affiliations, has been extracted and disambiguated in separate tables. In addition, the disambiguated authors have unique identifiers AIDs.

Table 1. A description of main tables in the downloaded PubMed dataset provided in (Xu et al., 2020).

| Table | # Of rows | # Of distinct entities | Description |
|---|---|---|---|
| A01_Articles | 31,928,777 | 31,926,861 | A table containing PubMed articles' bibliographic information. |
| A02_AuthorList | 131,446,038 | 18,519,492 | A table containing PubMed authors and their unique identifier. |
| B10_BERN_Main | 295,921,671 | 20,136,150 | A table containing all types of extracted bio-entities by BioBERT are used in both building the Knowledge Graph and as an index. |
| C03_Affiliation_Merge | 62,015,712 | 9,502,394 | Table containing affiliations and them extracted fine-grained items. |
| C05_NIH_PubMed | 22,946,601 | 116,530 | A table containing projects from NIH ExPORTER and mapping relation between PI_ID, PMID, and AND_ID. |
| C04_ReferenceList | 633,401,975 | 23,856,949 | A table containing reference relations between PMID and reference PMID. It was extracted from the Web of Sciences. |

Table 1 provides statistics and a description of the PubMed relational database for essential tables. The original dataset contains 27 tables linked by PMID. Here we extract metadata from 7 tables. In addition, we do not use 31 million articles for our dataset. Instead, we choose a subset of articles that have been submitted to journals between the dates of 2/1/2019 and 2/3/2019. This subset of the articles yielded 2696 articles when queried on PubMed. We then use the 2696 articles to extract a first-order citation network from the table C04_ReferenceList. The citation network yielded 100456 articles. Finally, for the 100456 articles, we extracted the rest of the metadata from the tables listed above, which will be described later.

## 3.2 Indexing

Indexing refers to extracting unique vocabulary from the corpus of documents and creating an index of the terms in each dataset article. In our case, the table named B10_BERN_Main represents the names of drugs, genes, diseases, species extracted using named entity recognition using the biomedical deep learning language model BioBERT (Lee et al., 2020) in the dataset presented in (Xu et al., 2020), which acts as an index in addition to being part of the knowledge graph that we will describe its extraction later in the following subsection. In addition, the index will be used to match input user queries and expansion and create query vectors. More formally, each article $p \in P$ will contain a set of biological entity mentions $m \in M$. Each mention is part of a set of mentions that distinguish each unique biological entity $b \in B$ where $M' \subseteq M$ and $b \rightarrow |M'|$. In addition, each unique biological entity has a type that can be one of 4 types {drug, disease, gene, species} where $b(t) \in B(T)$ and $T = \{drug, disease, gene, species\}$ . Hence the relationship becomes $p(b(t)) \in P(B(T)) \forall b \rightarrow |M'|$ . Note that we only use extracted biological entities from the text of each article to index our corpus of articles instead of using

MeSH terms or UMLS (Bodenreider, 2004) vocabulary, which is considered a standard approach in work that has been done before in biomedical information retrieval and text mining.

### 3.3 Knowledge graph extraction

Knowledge graph extraction converts the relational database of the PubMed metadata to a knowledge graph of interconnected entities, as shown in figure 2.

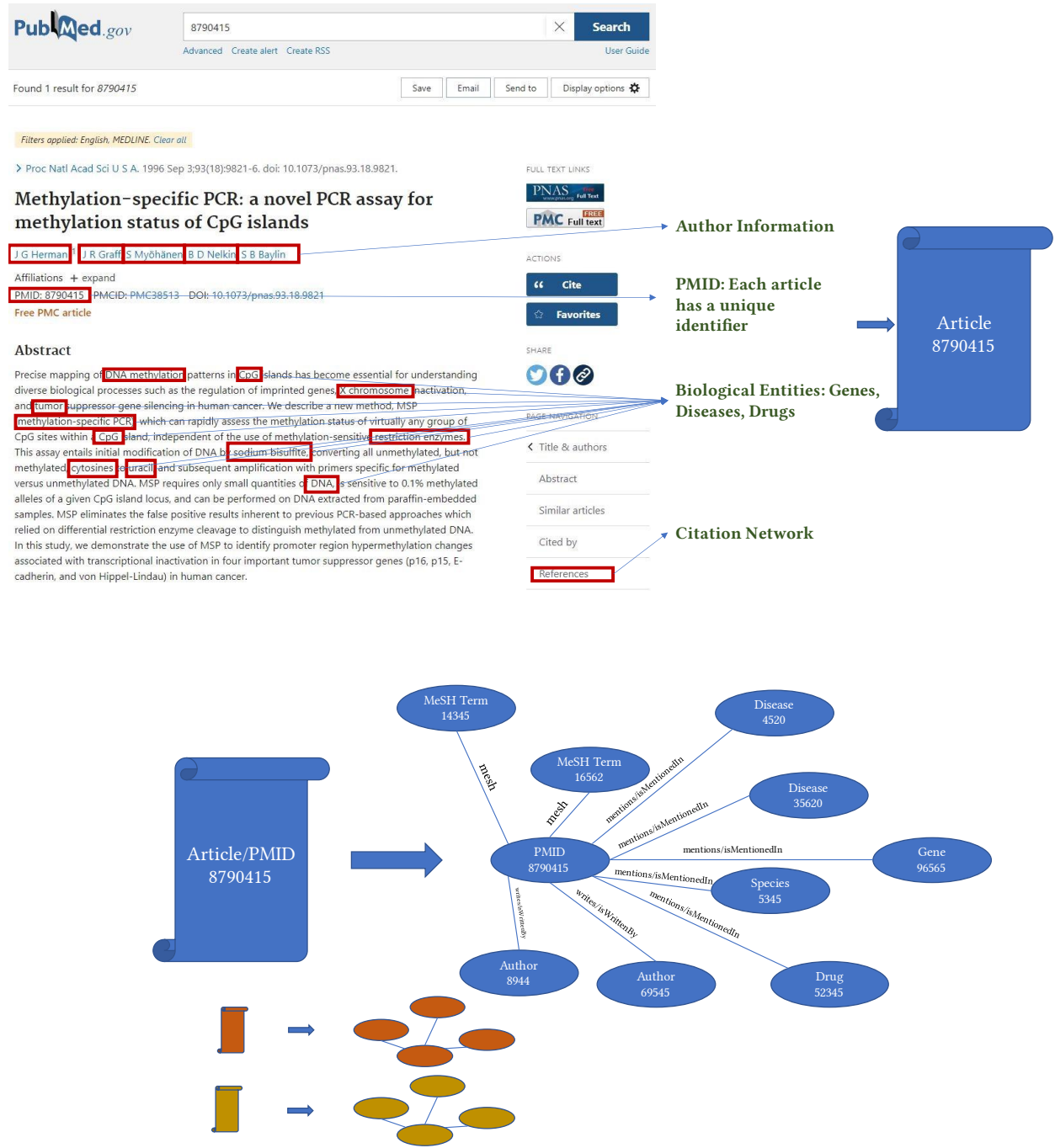

Figure 2. The image represents how each article is converted to a concept graph or a smaller knowledge graph.

For each article, we first extract all author names, names of drugs, genes, proteins, diseases, and species in addition to related MeSH terms and Chemical Substances terms from the tables described above. Then, the unique identifiers representing each entity create the knowledge graph. As described before, knowledge graphs are represented as a list of triples. For example, in our case, when we extract an author name for an article from the metadata database, we represent that information as <"article/pmid/86509", "isWrittenBy/wrote," "author/aid/6754">. Similarly, when we extract a drug name from an article, that information is represented as <"bioentity/drug/1256", "isMentionedIn/mentions," "article/pmid/78456">. In addition, if an NIH grant or project funded an article, that information will be represented as <" article/pmid/5678", "isFundedBy/funds," "nih_project/project_id/4123">. Note that the relationships are represented equally as the data in this knowledge graph model compared with a relational model.

Accordingly, each article and associated metadata will be represented as a mini knowledge graph or a concept graph, as shown in figure 2. Those mini knowledge graphs or concept graphs could be seen as subgraphs of a larger encompassing knowledge graph. In our case, we link all the subgraphs in two ways. First, we use the citation network provided in table C04_ReferenceList, representing extracted citation information from PubMed and Web of Science. The citation network provides the set of edges necessary to link most of the articles using the relationship "isCitedBy/cites." For example, two articles will be linked and represented in the knowledge graph as a triple <"article/pmid/652148", "isCitedBy/cites," "article/pmid/415923">. Second, since the authors and the names of drugs, diseases, genes, and proteins are disambiguated and unique, if an author appears with multiple names across several articles, all the names they appeared with will have the same author identifier number. Similarly, they will have the same unique identifier if they appear with different names such as Aspirin and NSAID for drugs, proteins, genes, and species. Moreover, since for each article, we create the mini knowledge graph using a unique identifier. The linked knowledge graph will also be semantically linked because an author will appear in multiple articles, a drug name in multiple articles, and the citation network links all articles together. The final knowledge graph will be a semantically linked network representing articles, authors, NIH grants, drugs, diseases, genes. Extracting a knowledge graph dataset as described above for the whole corpus of articles in PubMed is a daunting task, so to have a proof of concept, we extract only a small subset of articles with their citation information. Knowledge graph extraction can be formalized by seeing each subject and object in the extracted triples $< v_i, r_k, v_j >$ as nodes in a knowledge graph $v(l) \in V(L)$ where each node has a type $l \in L = \{article, author, gene/protein, drug, disease, species, nih\_project, mesh\_term, chemical\_substance\}$. Edges in the knowledge graph are equivalent to verbs or predicates in the triple representation, as shown in figure 3. Each edge $e(k) \in E(K)$ has a type $k \in K = \{isCitedBy/cites, isMentionedIn/mentions, isFundedBy/funds, mesh, isRelatedTo/relates\}$. Hence the triple relationship can be reformalized to $G = (V, E)$.

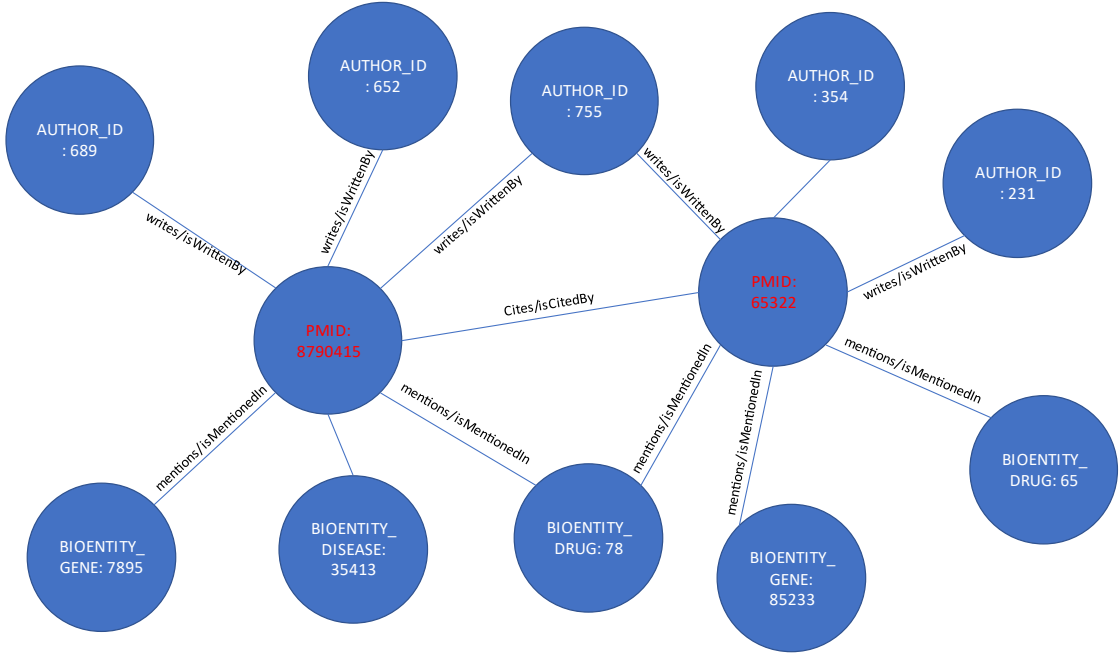

Figure 3. A part of the extracted knowledge graph

### 3.4 Knowledge graph embedding

As mentioned before, knowledge graph embedding models can be inductive as in learned from the structure of the graph itself (Perozzi et al., 2014) (Tang et al., 2015) (Grover & Leskovec, 2016) or transductive by forcing a scoring function to evaluate the plausibility of the triples in the knowledge graph (Lin et al., 2015) (Bordes et al., 2013). It is worth mentioning that knowledge graph embedding techniques are part of the larger field of graph representation learning, including GNNs. More knowledge graph embedding and transductive methods can be found in (Q. Wang et al., 2017). Knowledge graph embedding aims to learn a set of feature vectors for each node or entity in the knowledge graph. The feature vector needs to encode the structure of the graph. More formally, for the graph $G = (V, E)$ a matrix $X \in \mathbb{R}^d$ is learned via the function $f: v \in V \rightarrow \mathbb{R}^d$. One of the constraints on the learned embedding matrix is that it can be decomposed to $X = Z_v^T Z_u$ so that $X$ preserves the similarity between its component matrices where $v \in V$ and $u \in V$ and $Z_v = X^T$ and $Z_u = X$. Preserving the similarity is learned through predicting the probabilities of co-occurrence between 2 nodes in the same neighborhood within a specific context window C after sampling the graph using a random walk strategy to a size of a corpus sampled nodes, T.

$$P(v_1, v_2, v_3, \dots, v_t) = \frac{1}{T} \sum_{T}^{t=1} \sum_{-c \leq j \leq c, j=0} \log P(v_{t+j} | v_t)$$

Where $c \in C$ and $t \in T$. $\{v_1, v_2, v_3, \dots, v_t\}$ is sampled from the first order neighborhood N of a randomly chosen node $v_i$. To train matrix X, we approximate the probability $P(v_1, v_2, v_3, \dots, v_t)$ over positively and negatively sampled and labeled nodes using a sliding window on the sampled chains of nodes from the graph. Nodes within the context window are labeled one, while nodes outside the context window are labeled 0. A sigmoid function is then used to normalize the parameters of the matrix $X$. A backpropagation phase then takes place to optimize the loss function:

$$J_t(\theta) = \log \sigma(u_0^T v_c) + \sum_{j = P(V)} \log \sigma(-u_j^T v_c)$$

Where $u$ and $v \in V$ and $u_i$ and $v_i$ are row vectors $\in X$. The previously described algorithm is the Skip-gram model and was introduced in (Mikolov, Sutskever, et al., 2013). To extract knowledge graph embedding representations, we use Node2vec, the algorithm described in (Grover & Leskovec, 2016). Node2vec performs a modified version of the random walk strategy (Perozzi et al., 2014), including parameters p and q to control the sampling strategy. The p parameter controls the likelihood of the walk revisiting a node. The q parameter controls whether the search is constrained locally or globally. Given q > 1 and a random walk on an initial node, the random walk samples nodes closer to the initial node as in Breadth-First Search. Whereas q < 1, random walk samples nodes further from the initial node like a Depth First Search. This customizability in search behavior allows the random walker to capture diverse structural and topological properties within the graph. The sampling strategy builds a corpus of walks starting from each node. A Skip-gram model trains on this corpus to generate a unique embedding vector for each node in the knowledge graph as described above. Once the model finishes training, we get an embedding vector of size d for each node regardless of its type, whether an article, author, drug, disease, gene, NIH project, or MeSH term.

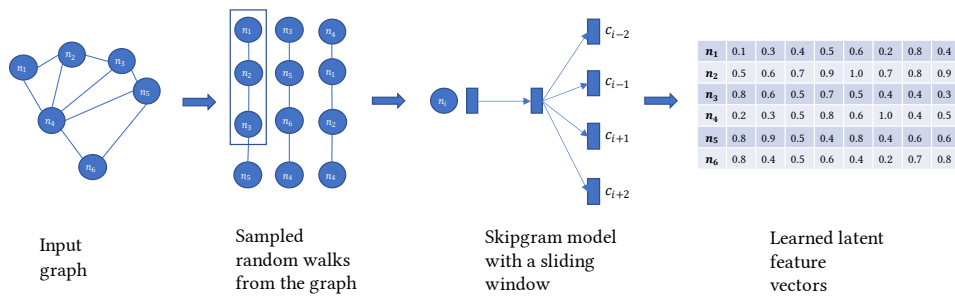

Figure 4. A representation of the Skip-gram model.

### 3.5 Article embedding

As mentioned before, our goal is to build a backend knowledge graph-based embedding model used by a front-end search engine to rank articles relevant to specific user queries. This step uses a pooling operation averaging all the node embedding vectors of all types of nodes connected to each article node in its first-order neighborhood. We created the article embedding model in 2 stages. First, we performed the pooling operation of averaging all the nodes of the articles as described before mentioned in the citation network, which gives us 100456 articles. Next, we did a second pooling operation where we averaged the first-order neighbors of articles for the 2696 articles we intend to search.

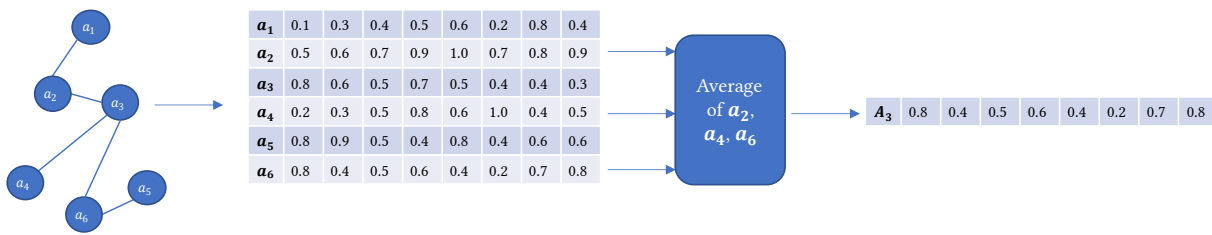

Figure 5. How article embeddings are generated.

In figure 5, the graph on the left is our knowledge graph, where we only have article nodes. For example, suppose we want to calculate the embedding for article a3, one of the 2696 articles, but it is also connected to other article nodes in the graph. So we average all the embedding vectors of the neighboring articles only, that is, a2, a4, and a6, and the resultant vector will be the one representing a3.

### 3.6 Query tokenizer

This module acts as an interface with the user. It takes user queries and parses them. The input queries are assumed to be in English and are tokenized by splitting over white spaces after removing punctuation, stop words and verbs. For example, a query like "show me articles on depression and type 2 diabetes" after tokenization it will be reduced to ["articles," "depression," "type," "2," "diabetes"]. The output keywords will be passed to the query expansion module. Note that the assumption here is that the query should include keywords in the index.

### 3.7 Query matcher

The list of extracted keywords is then expanded using a sliding window of sizes 2, 3, and 4. The sliding window's function captures multiple tokens from the initial keyword list. It slides over the list of keywords and expands it. For example, our list of keywords ["articles," "depression," "type," "2," "diabetes"] will be expanded to ["articles," "depression," "type," "2," "diabetes," "articles depression," "depression type," "type 2", "2 diabetes", "articles depression type," "depression type 2", "type 2 diabetes"]. The expanded list of keywords is then matched using a Levenshtein string distance comparator to the index. The index contains all the extracted biological entities from the articles and their unique identifiers and locations. For the matched mentions in each article in the index, each biological entity's unique identifier will be extracted and passed to the next step. The system exits if the keywords are not found in the index.

### 3.8 Query embedding

This step aims to find all the nodes in the knowledge graph with the same identifiers as the identifiers returned by the query matcher. After identifying the nodes, their corresponding learned embedding vectors from the knowledge graph embedding step is extracted. All the vectors are averaged to a single vector in a pooling operation like figure 5. The single vector becomes our query embedding vector.

### 3.9 Cosine distance and ranked results

In a Euclidean space, the cosine of angle θ between two vectors A and B is determined using the relationship:

$$\text{similarity } = \cos(\theta) = \frac{AB}{\|A\|\|B\|}$$

Since our knowledge graph has been embedded in Euclidean space, the similarity between two nodes is equivalent to the cosine of the angle between the two vectors representing the two nodes. So at this point, we have a query vector and a set of article vectors a simple operation between the query vector and the article vectors would yield the list of articles that are relevant to the query vector, and when sorted by the cosine score, the list of articles will be presented as ranked retrieved articles.

## 4 Evaluation

### 4.1 Dataset

As mentioned before, we extracted a proof of concept dataset from the PubMed database described in (Xu et al., 2020) and available at (http://er.tacc.utexas.edu/datasets/ped). The database contains 3190000 articles indexed from the year 1781 to December 2020. We extracted our target dataset of 2696 articles submitted to journals between 02/01/2019 and 02/03/2019. We came about those dates by examining the number of articles that have been submitted to journals for each month in the past five years in PubMed. We then chose the month with the least number of articles submitted, February 2019. Still, the dataset at that point was too large. Note that we include extracted articles, but we also query the reference table to extract the first order citations of each article, so the number grows exponentially. Accordingly, we kept reducing the number of days where articles were submitted to their journals until we got a reasonable size dataset. The dataset was extracted by first querying the PubMed online search engine (https://pubmed.ncbi.nlm.nih.gov/) for the articles that were submitted to their journals each month for each year since 2019:

(((("2019/month/01"[Date - Completion]: "2019/month/30"[Date - Completion]))))

Then the month with the least number of completed and submitted articles was chosen across all years. Then we adjusted to choose only three days since the size of the yielded citation network would have been beyond the scope of this study. We then settled for the dates mentioned above and queried PubMed with the query which yielded 2696 articles:

(((("2019/02/01"[Date - Completion]: "2019/02/03"[Date - Completion]))))

We then extracted the PMIDs of those articles. The extracted PMIDs were used to query the downloaded PubMed database to extract all the necessary metadata for each article. We first extracted the citation network of the 2696 articles, which yielded 100456 articles, including the 2696 articles. For the 100450 articles, all the metadata has been extracted, including author information, MeSH terms, Substances, NIH project involvement, extracted drug, disease, and protein names, and citation network from the tables described in section 3.2. The extracted metadata was used to create the knowledge graph described in section 3.3. The final knowledge graph is a multi-undirected graph with the following description in table 2. The total nodes in the graph were 578453, representing nine types of entities; authors, articles, NIH projects, MeSH terms, registered chemical substances, diseases, drugs, genes, and species. Most of those nodes were author nodes, followed by article nodes, then several NIH projects, MeSH term nodes, and extracted biological entities. Note that what defines a node in a graph is its identifier. Each node in the knowledge graph is identified by its original identifier concatenated to its type with a slash. For authors, identifiers are Author IDs (AIDs) in the database, PMIDs identify articles, Project IDs identify NIH projects, Header IDs identify MeSH terms, and extracted biological entities are identified by their unique Entity ID assigned by BioBERT in the original paper (Xu et al., 2020). For example, an article node will appear in the knowledge graph "article/pmid/652148". On the other hand, edges in the knowledge graph are identified by their edge type. Here we identify 9 relationships represented with edge labels, as shown in table 2.

Table 2. The description of node and edge types in the extracted knowledge graph.

| Node/Edge type | Number of nodes in the graph |
|---|---|
| # **Nodes** | 578453 |
| # Author | 393864 |
| # Article | 100456 |
| # NIH Projects | 27109 |
| # MeSH Terms | 20015 |
| # Chemical Substances | 9686 |
| # Disease | 9594 |
| # Drug | 8762 |
| # Gene | 6094 |
| # Species | 2873 |
| **# Edges** | **2226999** |
| # Article-relatedTo-MeSHTerm | 1049789 |
| # Article-writtenBy-Author | 596340 |
| # Article-mentions-Disease | 176516 |
| # Article-mentions-Drug | 108435 |
| # Article-cites-Article | 104138 |
| # Article-mentions-Species | 70694 |
| # Article-mentions-Gene | 56337 |
| # Article-isFundedBy-NIHProject | 54751 |
| # Article-relatedTo-Substances | 9999 |

## 4.2 Experimental setup

We then trained the resultant knowledge graph to extract node embedding vectors using a Node2vec (Grover & Leskovec, 2016) approach implemented using Python 3.8 and the library Stellargraph (Data61, 2018). The algorithm first runs a biased random walk sampling algorithm on the graph to sample chains of nodes using the breadth-first bias parameter q=0.5 and the depth-first bias parameter p=2 with a walk length of 50 and 5 walks per node. The sampled corpus of node walks is then used to train a Skip-gram model as described in section 3.4. Next, we tuned the Skipgam model, and the final model was trained using the vector size 128, context window size 5, and the number of negative samples was 7. The model was trained on a Windows PC with an Intel i7 processor and 32 GB of RAM. We also implemented and trained a TF-IDF model on our corpus of 100456 articles then extracted the TFIDF vectors for the 2696 target articles to compare against our method. With the help of the Python library Gensim (Rehurek & Sojka, 2011), we first extracted a dictionary of unique tokens in the corpus then trained a Bag of Words model. The Bag of Words model was then used to train the TF-IDF model, yielding a vectorized document matrix and unique vocabulary.

We evaluated MedGraph to assess the quality of our knowledge graph embedding based relevance ranking against PubMed's BestMatch algorithm as ground truth. We extracted a set of 15 queries from PubMed, and we applied the search on the articles that have been completed between the dates of 2/1/2019 and 2/3/2019. The 15 queries were chosen randomly from the extracted index of biological entities as described in section 3.2. They contained the names of diseases and drugs, as shown in table 3. For example, for the query "type 2 diabetes," we use the following query to search PubMed then download the resultant PMIDs of the ranked articles.

((((("2019/02/01"[Date - Completion] : "2019/02/03"[Date - Completion])))) AND (type 2 diabetes[Text Word])

Table 3. A description of the queries we used to evaluate the system against PubMed's BestMatch ranked results were used as a ground truth.

| Query ID | Text | # Relevant Documents | # Tokens |
|---|---|---|---|
| 1 | **alcohol** | 37 | 1 |
| 2 | **amino acids** | 11 | 2 |
| 3 | **bacterial infections** | 6 | 2 |
| 4 | **basal cell carcinoma** | 3 | 3 |
| 5 | **bipolar disorder** | 10 | 2 |
| 6 | **cancer** | 320 | 1 |
| 7 | **diabetes** | 59 | 1 |
| 8 | **hepatitis c virus** | 3 | 3 |
| 9 | **histamine** | 2 | 1 |

| 10 | **insulin** | 25 | 1 |
| 11 | **loss of muscle strength** | 1 | 4 |
| 12 | **pediatric cancer** | 1 | 2 |
| 13 | **trauma** | 22 | 1 |
| 14 | **type 2 diabetes** | 22 | 3 |
| 15 | **urinary tract infection** | 5 | 3 |

Then for each query, we rank the articles based on the cosine distance metric by comparing the query vector to the article vectors described in section 3. We then prune the list of the resultant ranked retrieved articles by K. That means we choose the top K elements of the ranked retrieved articles from MedGraph. Then we compute the number of relevant articles, the number of retrieved articles, and the number of relevant articles retrieved. We then compute precision, recall, and f1-score. Precision is the number of relevant articles retrieved over the total number of relevant articles. The recall is the number of relevant articles retrieved over the total number of retrieved articles. Moreover, F1-score is the harmonic mean of both precision and recall.

We also compute the Mean Average Precision (MAP) across queries (Aslam & Yilmaz, 2006). MAP is a widely used metric in information retrieval to evaluate search engines. It focuses on precision since recall can be misleading in some cases. To compute MAP, we first compute the average precision for each query. That is done by finding each retrieved article in the ground truth and for top K. Then computing precision at each article in the retrieved articles. That is followed by averaging the precision values across all retrieved articles K. Then averaging across all the queries.

### 4.3 Results

Table 4 presents the results on the four metrics we described in section 4.2. We ran 12 levels of K for both our method MedGraph and the standard TF-IDF (Ramos & others, 2003) approach for ranking relevant documents. Our results indicate that MedGraph has outperformed TF-IDF on the PubMed BestMatch dataset at various levels of K and across all queries and metrics. The only exception is that MAP at higher K levels was higher for TF-IDF. That might explain why TF-IDF returns more relevant documents but does not rank them higher, while MedGraph might retrieve less relevant documents more semantically related and ranked closely. In addition, both precision and recall for MedGraph were consistently higher with the recall increased exponentially with higher K, and precision decreased exponentially with higher K levels, as demonstrated in figure 6.

Table 4. Results averaged across the 15 queries

| Metric | Method | K=1 | K=2 | K=5 | K=10 | K=25 | K=50 | K=75 | K=100 | K=150 | K=250 | K=500 | K=1000 |
|---|---|---|---|---|---|---|---|---|---|---|---|---|---|
| Recall | TFIDF | 0.053 | 0.062 | 0.078 | 0.113 | 0.177 | 0.197 | 0.202 | 0.207 | 0.217 | 0.22 | 0.22 | 0.22 |
| | MedGraph | **0.227** | **0.245** | **0.297** | **0.392** | **0.545** | **0.646** | **0.693** | **0.726** | **0.749** | **0.846** | **0.931** | **0.976** |
| Precision | TFIDF | 0.467 | 0.4 | 0.307 | 0.293 | 0.248 | 0.171 | 0.136 | 0.117 | 0.1 | 0.064 | 0.032 | 0.016 |
| | MedGraph | **0.867** | **0.667** | **0.507** | **0.453** | **0.336** | **0.248** | **0.202** | **0.172** | **0.134** | **0.103** | **0.064** | **0.034** |
| F1-Score | TFIDF | 0.081 | 0.083 | 0.09 | 0.122 | 0.161 | 0.138 | 0.118 | 0.107 | 0.097 | 0.074 | 0.046 | 0.026 |
| | MedGraph | **0.279** | **0.245** | **0.235** | **0.276** | **0.282** | **0.253** | **0.221** | **0.197** | **0.161** | **0.134** | **0.095** | **0.056** |
| MAP | TFIDF | 0.467 | 0.383 | 0.272 | 0.221 | 0.184 | 0.177 | 0.175 | 0.174 | 0.174 | 0.173 | 0.173 | 0.173 |
| | MedGraph | **0.867** | **0.55** | **0.284** | 0.168 | 0.077 | 0.041 | 0.028 | 0.021 | 0.014 | 0.009 | 0.004 | 0.002 |

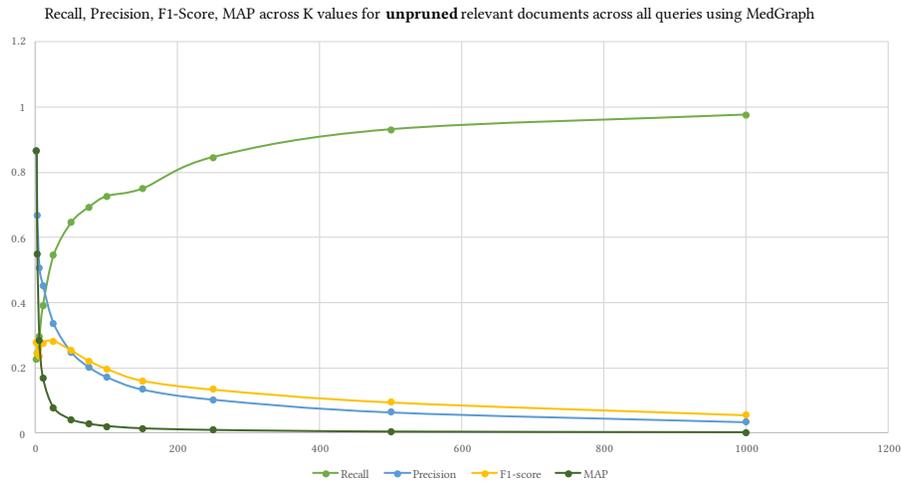

Recall, Precision, F1-Score, MAP across K values for **unpruned** relevant documents across all queries using MedGraph

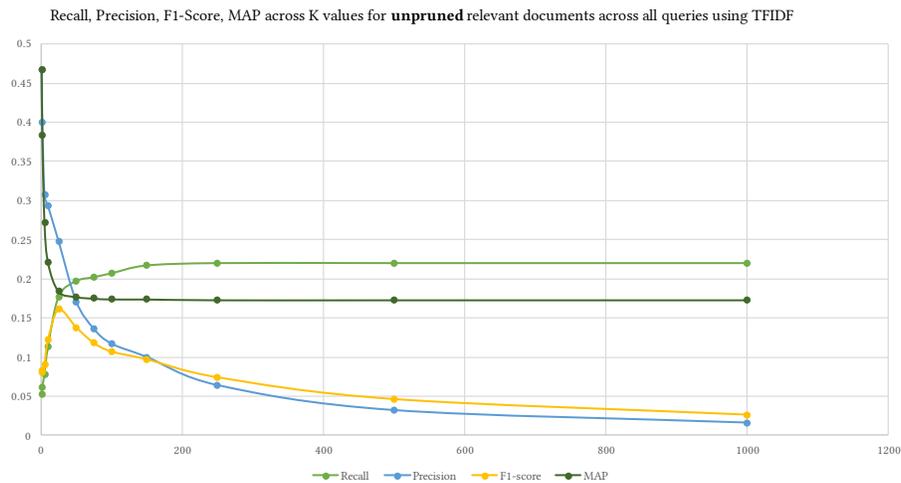

Recall, Precision, F1-Score, MAP across K values for **unpruned** relevant documents across all queries using TFIDF

Figure 6. MedGraph versus TF-IDF on all four metrics

MedGraph had higher MAP and F1 Scores across all K levels due to its higher recall and precision. The highest difference between MedGraph and TF-IDF was at K=1, indicating that the first document in the retrieved documents almost always existed in the ground truth dataset. However, recall was the lowest because most of the relevant documents did not exist in the first position in the retrieved documents. Of course, as we increase K, the recall increases, indicating that most of the relevant documents in the ground truth appeared in the retrieved documents. At K = 10, MedGraph started underperforming on MAP while TF-IDF stayed consistent at higher K levels. That is because MedGraph ranks a small number of the relevant documents highly, while a large number of the documents do not appear in MedGraph. The documents that appear in the retrieved documents are ranked closely and higher due to the semantic nature of the algorithm, while the documents that are not closely ranked and in the top are usually ranked lower and tend to be spread out.

Alternatively, in other words, MedGraph produces relevant articles that are closely ranked together due to the semantic nature of the algorithm, while TF-IDF produces almost the same number of relevant articles but are not ranked closer together. Finally, we computed the four metrics by pruning the top K ground truth results relevant documents from BestMatch.

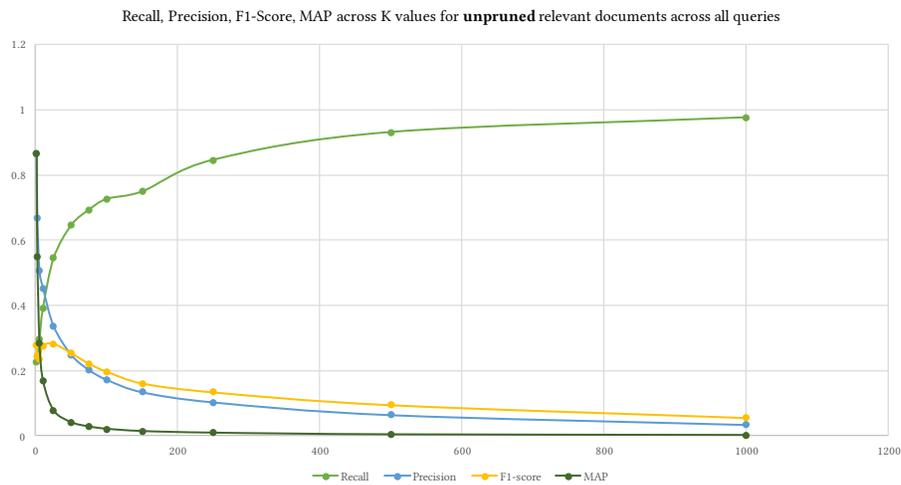

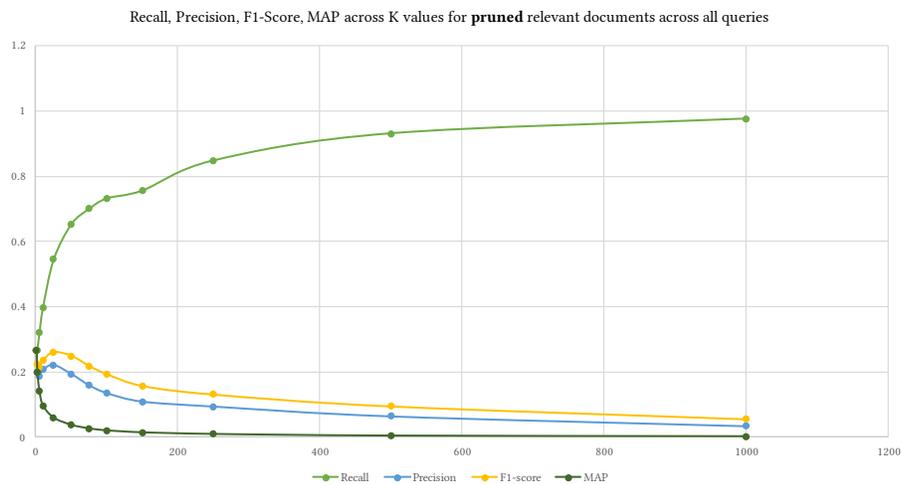

Figure 7. The difference between pruned and unpruned retrieved results

We used the same K levels provided to prune the retrieved and relevant results. Figure 7 shows the difference between pruning the relevant ground truth articles and not pruning them. The values of recall and F1-Score do not differ between both approaches, yet precision and MAP are higher when the relevant documents are not pruned using K. Pruning perhaps provides a mechanism to control the ground truth dataset. We do not know how exactly BestMatch ranked it. The returned BestMatch articles from PubMed have different retrieved articles without explanation, as shown in table 3. Hence pruning might make sense in some cases depending on the evaluation dataset.

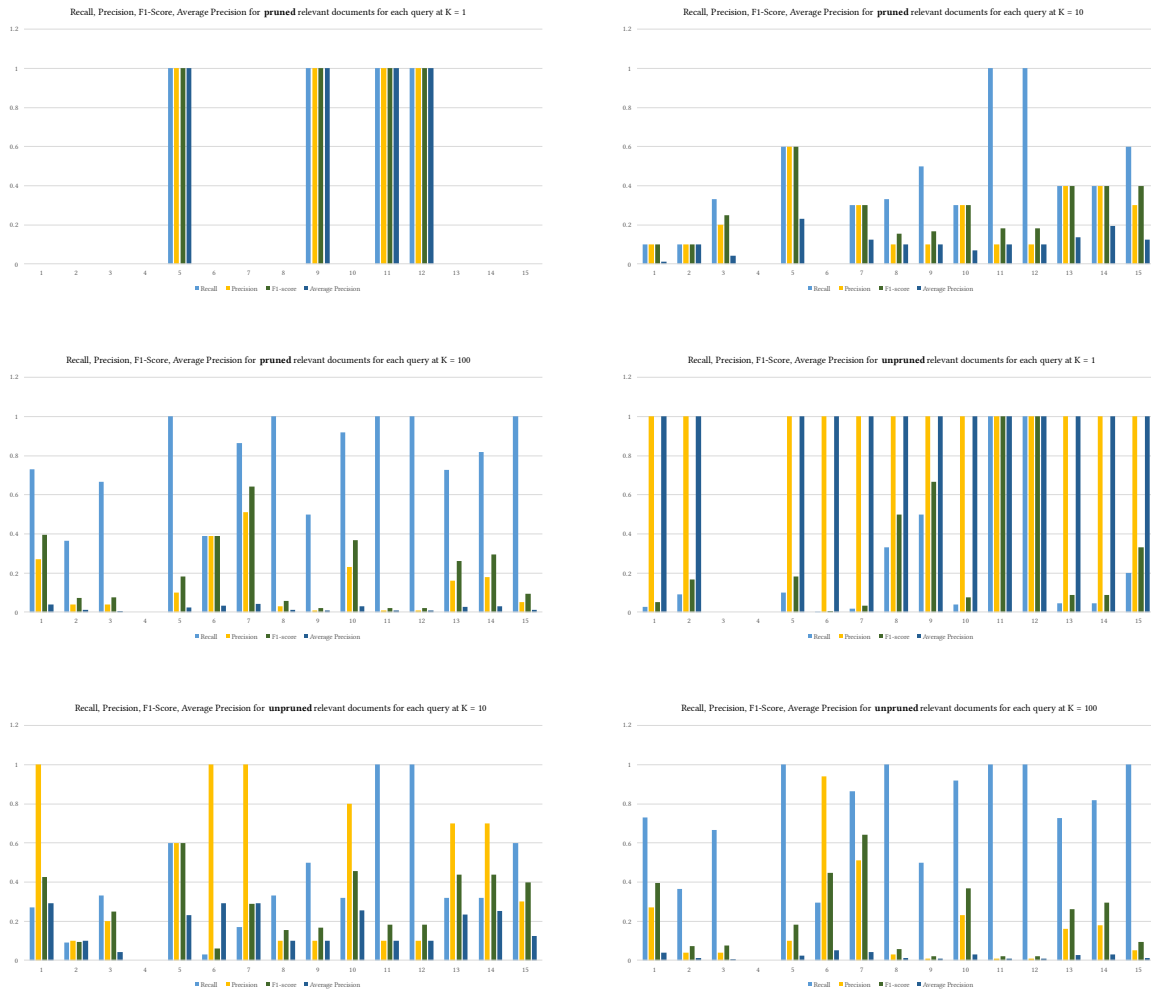

Figure 8. The four metrics across various levels of K over the 15 queries. Upper is pruned, and lower is unpruned relevant documents.

That is also seen in figure 8, where precision was much higher across queries with unpruned relevant results, indicating that MedGraph retrieved almost all of the relevant results compared with BestMatch.

## 5 Discussion

This work provided evidence of the utility and efficiency of knowledge graph-based methods in information retrieval, especially in the biomedical field. We highlighted the need for more techniques that rely on semantic understanding of queries and datasets to aid in automated knowledge discovery and information organization. Knowledge graphs have been around for a while, yet they have not been fully utilized in search engines. Approaches such as BestMatch for PubMed are very efficient but do not understand semantics and are trained on user query logs that might change over time, requiring retraining. Traditional TF-IDF approaches do not rely on semantics and are almost outperformed by newly developed methods like ours. The results also indicated that MAP alone is not enough as an evaluation metric. The ranking is usually evaluated using A/B testing approaches involving user studies and metrics that would include users ranking relevance by hand then computing metrics such as Normalized Discounted Cumulative Gain (Busa-Fekete et al., 2012). Precision as a metric is very informative as evaluating how many relevant articles were retrieved and, in our case, MedGraph. It highlighted its superiority. Nevertheless, metrics such as recall can be misleading. For example, if the system only retrieved one document, but that document is in the relevant documents no matter the rank, then recall shall be 100%. Precision acts as a self-assessment on the retrieved articles by MedGraph because it compares the numbers of retrieved relevant articles to the number of retrieved articles regardless of the number of relevant articles.

In our future work, we plan to conduct a user study where each user, typically a biomedical researcher or a medical student, will be invited and asked to rank documents based on specific queries. We will create our ground-truth dataset instead of relying on BestMatch as our ground truth. We also plan to expand the scope to extract a knowledge graph from the entire dataset of 30 million articles described in (Xu et al., 2020) and compare our model with BestMatch and TFIDF using our ground truth. In addition, we plan to experiment with various other knowledge graph embedding models (Q. Wang et al., 2017). For example, Node2vec represents a basic model incapable of encoding heterogeneity in knowledge graphs. Heterogeneity refers to a knowledge graph having more than one type of node and more than one type of edge or relationship. Hence, we will experiment with models capable of capturing more semantics in the training of node embeddings, expanding our query matching capabilities to include more than four tokens, and handling out-of-context queries. In addition, we plan to have even more metadata nodes in our knowledge graph with the potential of enriching the knowledge graph with other semantic datasets such as Chem2Bio2RDF (Chen et al., 2010).

## 6 Conclusion

In this work, we presented a proof-of-concept method to build a semantic search engine for the biomedical literature indexed in PubMed named MedGraph. We showed that our method is superior to more traditional approaches in relevance ranking and provided evidence that semantic methods in information retrieval are more needed than ever. Furthermore, we performed a complete evaluation using various metrics on our approach using PubMed's BestMatch as a ground truth. We also presented an innovative way of converting relational databases to knowledge graphs. In the future, we hope to expand this work and provide a fully working model and system accessible by researchers to provide better ways to discover knowledge and advance science.